\documentclass[aps,prd,twocolumn,superscriptaddress,groupedaddress,nofootinbib]{revtex4}  
\usepackage{graphicx}  
\usepackage{dcolumn}   
\usepackage{bm}        
\usepackage{amssymb}   
\usepackage{color}
\begin{document}

\title{Bopp-Podolsky black holes and the no-hair theorem}

\author{R. R. Cuzinatto}
\email{rodrigo.cuzinatto@unifal-mg.edu.br}
\affiliation{Department of Physics, McGill University, Ernest Rutherford Physics Building, 3600 University Street, H3A 2T8, Montreal, Quebec, Canada. }
\affiliation{Instituto de Ci\^{e}ncia e Tecnologia, Universidade Federal de Alfenas.\\Rod. Jos\'{e} Aur\'{e}lio Vilela (BR 267), Km 533, n. 11999, CEP 37701-970, \\Po\c{c}os de Caldas, MG, Brazil. }

\author{C. A. M. de Melo}
\email{cassius.anderson@gmail.com}
\affiliation{Instituto de Ci\^{e}ncia e Tecnologia, Universidade Federal de Alfenas.\\Rod. Jos\'{e} Aur\'{e}lio Vilela (BR 267), Km 533, n. 11999, CEP 37701-970, \\Po\c{c}os de Caldas, MG, Brazil. }
\affiliation{Instituto de F\'{\i}sica Te\'{o}rica, Universidade Estadual Paulista.\\Rua Bento Teobaldo Ferraz 271 Bloco II, P.O. Box 70532-2, CEP 01156-970, \\S\~{a}o Paulo, SP, Brazil. }

\author{L. G. Medeiros}
\email{leogmedeiros@ect.ufrn.br}
\affiliation{Instituto de F\'{\i}sica Te\'{o}rica, Universidade Estadual Paulista.\\Rua Bento Teobaldo Ferraz 271 Bloco II, P.O. Box 70532-2, CEP 01156-970, \\S\~{a}o Paulo, SP, Brazil. }
\affiliation{Escola de Ci\^{e}ncia e Tecnologia, Universidade Federal do Rio Grande do Norte.\\Campus Universit\'{a}rio, s/n - Lagoa Nova, CEP 59078-970, \\Natal, RN, Brazil. }

\author{B. M. Pimentel}
\email{pimentel@ift.unesp.br}
\affiliation{Instituto de F\'{\i}sica Te\'{o}rica, Universidade Estadual Paulista.\\Rua Bento Teobaldo Ferraz 271 Bloco II, P.O. Box 70532-2, CEP 01156-970, \\S\~{a}o Paulo, SP, Brazil. }

\author{P. J. Pompeia}
\email{pompeia@ita.br}
\affiliation{Departamento de F\'{\i}sica, Instituto Tecnol\'{o}gico de Aeron\'{a}utica.\\Pra\c{c}a Mal. Eduardo Gomes 50, CEP 12228-900, \\S\~{a}o Jos\'{e} dos Campos, SP, Brazil. }


\bigskip
\date{\today}

\begin{abstract}
Bopp-Podolsky electrodynamics is generalized to curved space-times.
The equations of motion are written for the case of
static spherically symmetric black holes and
their exterior solutions are analyzed using Bekenstein's
method. It is shown the solutions
split-up into two parts, namely a non-homogeneous (asymptotically massless) regime and a
homogeneous (asymptotically massive) sector which is null
outside the event horizon. In addition, in the simplest
approach to Bopp-Podolsky black holes, the non-homogeneous
solutions are found to be Maxwell's solutions leading to a
Reissner-Nordstr\"om black hole. It is also demonstrated that the only
exterior solution consistent with the weak and null energy conditions is the
Maxwell's one. Thus, in light of energy conditions, it is
concluded that only Maxwell modes propagate outside the horizon and,
therefore, the no-hair theorem is satisfied in the case of
Bopp-Podolsky fields in spherically symmetric space-times.
\end{abstract}

\maketitle

\section{Introduction}

It was just one year after the proposal of General Relativity by Einstein in
1915 \cite{Einstein} that the first analytical solution
to his gravitational field equations was obtained.
Schwarzshild \cite{Schwarzschild} proposed a spherical symmetric solution for
the gravitational/metric field produced by a point mass. On the same year,
Reissner \cite{Reissner} proposed a solution to a charged
point mass, which was two years later reconsidered by Nordstr\"om
\cite{Nordstrom} in a spherical coordinate system known today as the
Reissner-Nordstr\"om solution. These solutions were the first ones predicting
the existence of event horizons for very compact objects. These works paved
the way for a whole new area of research in gravitation commonly known as
black hole (BH) physics.

The decades of 1960 and 1970 witnessed a boom of interest in this area. In
1963, Kerr \cite{Kerr} presented his solution for a spinning mass, a result
that was generalized by Newman two years later with the introduction of
electric charge to the rotating body \cite{Newman1,Newman2}. A
few years later, important developments related to the
interaction between matter and gravitational fields were achieved. In this
line of research, a relevant contribution was given by Israel \cite{Israel1}
in 1967, when he proposed the first version of the \emph{no-hair
theorem/conjecture} for the spherically symmetric black holes.
This result was soon extended to include rotating and charged BHs
\cite{Israel2,Carter} and a final version of this theorem states that an
exterior solution of a BH is completely characterized by its mass, electric
charge and angular momentum. All other features of particles
(\textquotedblleft\emph{hair}\textquotedblright) have no contribution for the
gravitational properties of the black hole once these particles are inside the
event horizon. This theorem has been demonstrated for many cases
\cite{Bekenstein1972,Teitelboin1972,Bekenstein1972 v2} and for different
theories of gravity \cite{Hawking1972,Mayo and Bek 1996,Faraoni and Sotiriou
2012} but several results suggest its validity is limited -- see
\cite{Zlosh 2005,Herdeiro and Radu 2015} and references therein.
As a matter of fact, haired solutions appear for complex Proca field
in the vicinity of spinning BHs \cite{Herdeiro and Radu 2016}; in addition,
Ref. \cite{East2017} presents the possibility super-radiance could amplify quantum effects
making these hairs potentially detectable by gravitational-wave observatories.

The case of a real Proca field is a particularly interesting case
studied by Bekenstein in \cite{Bekenstein1972}. He analyzed BHs
in the presence of a massive vector field
(henceforth called Proca black hole). This is a compelling
case which verifies the statement in the no-hair theorem.
Bekenstein builds an ingenious argument to show that the massive field
 cannot propagate outside the event horizon without
making use of an analytical explicit solution\footnote{Until now, there are only perturbative
\cite{Gott1984,Vuille2002} and numerical
\cite{Rosen1994,Obuk2000,Toussaint2000} solutions to the Proca
black hole.}. This way, no information about the mass of the Proca field can
be obtained by an observer outside the BH. This case is in contrast with the
massless (Maxwell) vectorial field, which is clearly known to propagate
outside the event horizon. So, at this point, one could ask: Why can Maxwell
field propagate outside the event horizon whilst Proca field
can not?

The two most significant differences between Proca and Maxwell fields
lay on the fact that the last is a massless gauge invariant
field while the former is a massive non-gauge invariant field. Thus, the
question above may be reformulated as: Can gauge invariance
and mass/massless property be the keystone for the difference
concerning field propagation between these two
cases? In order to answer this question, we will consider an
extension of Maxwell's theory proposed in the early 1940's by
Bopp and Podolsky {\cite{Bopp1940,Podolsky1942}}, known today as Bopp-Podolsky
electrodynamics, or shortly Podolsky electrodynamics. The Lagrangian of this
theory is characterized by possessing, in addition to the
usual Maxwell term, a term depending on the second derivative
of the gauge field, leading to fourth-order field equations. Although higher
order theories usually suffer from instabilities (ghost at the
quantum level), they can be avoided
in Podolsky's case by using the concept of Lagrange anchor
\cite{Kap2014}. Moreover, Podolsky electrodynamics presents
unique properties that make it worth analyzing. For instance,
this theory has been proven to be the only second order gauge theory for the
$U(1)$ group to preserve the linearity of the field equations
{\cite{AnnalsRoCaPJ}}. Also, the solution of the field equations shows
Podolsky field splits in two modes: a massive mode and a
massless one. These two properties set
Podolsky field as one of the most promising
candidates to properly address the question raised before. It
is important to emphasize that these properties of Podolsky electrodynamics
are valid in flat space-time. Thus, it is essential to verify
if they are still valid in curved space-time. Although some aspects of this
theory in curved space-time are found in the literature
\cite{Zayats2013,Zayats2016,Haghani2016,EPL2017}, these two properties will be
analyzed in some detail here.

Our intention in this paper is to understand the propagation of vector fields
outside the event horizon for Podolsky black holes. In particular, we are
interested in verifying if the no-hair theorem remains valid. For this, we will
analyze the properties of the electrostatic spherically symmetric solution of
Podolsky electrodynamics in curved space-time. In Section 2, we
will analyze how Podolsky electrodynamics is properly generalized to curved
space-time, considering its gauge invariance and linearity properties. Next, in
Sections 3 and 4, we shall investigate the properties of the
exterior solution using Bekenstein's approach. In Section 5, we reanalyze
these properties under the scrutiny of the null and weak energy conditions.
The final remarks are presented in Section 6.

\section{Podolsky Electrodynamics in curved space-time}

From a formal point of view, Podolsky electrodynamics
Lagrangian $\mathcal{L}_{m}$ is built under the following assumptions:

\begin{enumerate}
\item $\mathcal{L}_{m}$ must be invariant under Lorentz transformations (in
flat space-times);

\item $\mathcal{L}_{m}$ must be gauge invariant under $U(1)$
symmetry group, i.e. under a transformation of the type
$A_{\mu}\rightarrow A_{\mu}+\partial_{\mu}\alpha$;

\item $\mathcal{L}_{m}$ must be quadratic in the gauge field and its
derivatives resulting in linear field equations;

\item $\mathcal{L}_{m}$ is dependent on the gauge field and its first two derivatives.
\end{enumerate}

The authors of Ref. \cite{AnnalsRoCaPJ} followed the approach developed by R. Utiyama \cite{Utiyama1956} to show that
$\mathcal{L}_{m}$ is a combination of Maxwell Lagrangian and terms
of the form $\mathcal{L}_{m}\sim\partial_{\cdot}F^{\cdot\cdot}\partial_{\cdot}F^{\cdot\cdot}$
in Minkowski space-time, where the repeated symbol ``.''
indicates indices contraction. If we analyze all possible
contractions of indices, we verify that only three non-null and non trivially
equivalent terms remain:
\begin{eqnarray*}
\mathcal{L}_{m}^{(1)}  &  =\partial_{\beta}F^{\alpha\beta
}\partial_{\gamma}F_{\alpha}^{\text{ \ }\gamma}, \\
\mathcal{L}_{m}^{(2)}  &  =\partial_{\beta}F^{\alpha\gamma
}\partial_{\gamma}F_{\alpha}^{\text{ \ }\beta}, \\
\mathcal{L}_{m}^{(3)}  &  =\partial_{\beta}F^{\alpha\gamma
}\partial^{\beta}F_{\alpha\gamma}.
\end{eqnarray*}

It is not difficult to verify that $\mathcal{L}_{m}^{(2)}$
is equivalent to $\mathcal{L}_{m}^{(1)}$ (up to a surface
term) and that $\mathcal{L}_{m}^{(3)}$ can be obtained from
$\mathcal{L}_{m}^{(2)}$ when the Bianchi
identity \cite{AnnalsRoCaPJ} is taken into account. Therefore,
in flat space-time Podolsky electrodynamics is completely
described by the Lagrangian
\begin{equation}
\mathcal{L}_{m}^{flat}=-\frac{1}{4}F^{\alpha\beta}F_{\alpha\beta}+\frac{a^{2}%
}{2}\partial_{\beta}F^{\alpha\beta}\partial_{\gamma}F_{\alpha}^{\text{
\ }\gamma}. \label{L Pod flat}%
\end{equation}
Note that the positive sign in the second term allows
the factor $\frac{1}{a}$ to be interpreted as a mass parameter
\cite{IJMPA2011,BoninPimentel}. We consider the metric
signature $\left(  +,-,-,-\right)  $. Lagrangian $\mathcal{L}_{m}^{flat}$ was
proposed originally in the early 1940's by F. Bopp \cite{Bopp1940} and B.
Podolsky \cite{Podolsky1942}.

The next step is to generalize the approach presented in \cite{AnnalsRoCaPJ}
to curved space-times. For this end, in assumption 1 above, Lorentz invariance is
replaced by general covariance and a minimal coupling prescription is
considered, i.e. the following mappings apply: $\eta_{\mu\nu
}\rightarrow g_{\mu\nu},\,\partial_{\mu}\rightarrow\nabla_{\mu}$.
As a consequence, assumption 4 demands the Lagrangian
to be of the form $\mathcal{L}_{m}\left(  A,\nabla
A,\nabla\nabla A\right)  $. Next we impose the group symmetry condition
(assumption 2) under $U(1)$, which leads to:
\begin{eqnarray*}
\frac{\partial\mathcal{L}_{m}}{\partial A_{\mu}}\delta A_{\mu}&+&\frac
{\partial\mathcal{L}_{m}}{\partial\left(  \nabla_{\nu}A_{\mu}\right)  }%
\nabla_{\nu}\left(  \delta A_{\mu}\right) \\
&+&\frac{\partial\mathcal{L}_{m}%
}{\partial\left(  \nabla_{\lambda}\nabla_{\nu}A_{\mu}\right)  }\nabla
_{\lambda}\nabla_{\nu}\left(  \delta A_{\mu}\right)  =0,
\end{eqnarray*}
where $\delta A_{\mu}=\nabla_{\mu}\alpha$. By considering the
functional independence of $\alpha$ and its covariant derivatives\footnote{The requirement of functional independence of $\alpha$ and is ordinary derivatives leads to a system equivalent to Eqs.~(2-4).} we find:
\begin{eqnarray}
&&\frac{\partial\mathcal{L}_{m}}{\partial A_{\mu}}\nabla_{\mu}\alpha
=0,\label{1C}\\
&&\frac{\partial\mathcal{L}_{m}}{\partial\left(  \nabla_{\nu}A_{\mu}\right)
}\nabla_{\nu}\nabla_{\mu}\alpha =0,\label{2C}\\
&&\frac{\partial\mathcal{L}_{m}}{\partial\left(  \nabla_{\lambda}\nabla_{\nu
}A_{\mu}\right)  }\nabla_{\lambda}\nabla_{\nu}\nabla_{\mu}\alpha
=0.\label{3C}%
\end{eqnarray}

Analogously to what occurs in the case of flat space-time,
Eq.~(\ref{1C}) states $\mathcal{L}_{m}$ does not depend explicitly on
$A_{\mu}$. Then, from Eq.~(\ref{2C}) we see $\mathcal{L}_{m}$
depends on $\nabla_{\mu}A_{\nu}$ only through an antisymmetric combination,
given by
\begin{equation}
F_{\mu\nu}\equiv\nabla_{\mu}A_{\nu}-\nabla_{\nu}A_{\mu}, \label{F A}%
\end{equation}
and derivatives of $F_{\mu\nu}$. Note that the antisymmetry on $\nabla A$ was
established in view of the identity $\nabla_{\nu}\nabla_{\mu}\alpha
=\nabla_{\mu}\nabla_{\nu}\alpha$.

Now we implement assumptions $1$ and $3$ and verify that,
besides Maxwell term, only combinations of the form $\mathcal{L\sim}%
\nabla_{\cdot}F^{\cdot\cdot}\nabla_{\cdot}F^{\cdot\cdot}$ are allowed. As
before, only three non-null and non trivially equivalent terms remain
\begin{eqnarray*}
\mathcal{L}_{m}^{(1)}  &=&\nabla_{\beta}F^{\alpha\beta
}\nabla_{\gamma}F_{\alpha}^{\text{ \ }\gamma},\\
\mathcal{L}_{m}^{(2)}  &=&\nabla_{\beta}F^{\alpha\gamma
}\nabla_{\gamma}F_{\alpha}^{\text{ \ }\beta},\\
\mathcal{L}_{m}^{(3)}  &=&\nabla_{\beta}F^{\alpha\gamma
}\nabla^{\beta}F_{\alpha\gamma}.
\end{eqnarray*}
Direct substitution of these terms into Eq.~(\ref{3C})
shows this equation is satisfied for all terms.

Finally, we have to check if $\mathcal{L}_{m}^{(1)}$,
$\mathcal{L}_{m}^{(2)}$ and $\mathcal{L}_{m}%
^{(3)}$ are equivalent to each other. By using
the covariant version of Bianchi identity,
\begin{equation}
\nabla_{\beta}F_{\alpha\gamma}+\nabla_{\alpha}F_{\gamma\beta}+\nabla_{\gamma
}F_{\beta\alpha}=0,\label{Bianchi}%
\end{equation}
we show: $\mathcal{L}_{m}^{(3)}=2\mathcal{L}_{m}%
^{(2)}$. On the other hand, $\mathcal{L}%
_{m}^{(1)}$ and $\mathcal{L}_{m}^{(2)}$ are
related through\footnote{In this manuscript, the Riemann
tensor is defined by $R_{\text{ \ }\mu\nu\kappa}^{\sigma}\equiv\partial_{\nu
}\Gamma_{\mu\kappa}^{\sigma}-\partial_{\kappa}\Gamma_{\mu\nu}^{\sigma}%
+\Gamma_{\alpha\nu}^{\sigma}\Gamma_{\mu\kappa}^{\alpha}-\Gamma_{\alpha\kappa
}^{\sigma}\Gamma_{\mu\nu}^{\alpha}$.}
\begin{eqnarray}
\mathcal{L}_{m}^{(2)}  &=&\nabla_{\gamma}\left(  F_{\alpha
}^{\text{ \ }\beta}\nabla_{\beta}F^{\alpha\gamma}\right)  -\nabla_{\beta
}\left(  F_{\alpha}^{\text{ \ }\beta}\nabla_{\gamma}F^{\alpha\gamma}\right)
\nonumber\\
&+&R_{\sigma\beta}F_{\text{ \ }}^{\sigma\alpha}F_{\alpha}^{\text{ \ }\beta
}+R_{\alpha\sigma\beta\gamma}F^{\sigma\gamma}F^{\alpha\beta}+\mathcal{L}%
_{m}^{(1)}.\label{Nao equi L1 e L2}%
\end{eqnarray}
The first two terms on right-hand side are terms of the form $\nabla_{\mu
}V^{\mu}$, hence they constitute surface terms. However, the
non-commutativity of covariant derivatives imply the presence of two extra
terms consisting in non-minimal coupling between the field
strength and the Riemann tensor. These extra terms show
$\mathcal{L}_{m}^{(1)}$ and $\mathcal{L}_{m}^{(2)}$ are
not equivalent.

Thus, we conclude that the Podolsky electrodynamics in curved space-times
is given by the combination:
\begin{equation}
\mathcal{L}_{m}=-\frac{1}{4}F^{\alpha\beta}F_{\alpha\beta}+\frac{a^{2}}%
{2}\nabla_{\beta}F^{\alpha\beta}\nabla_{\gamma}F_{\alpha}^{\text{ \ }\gamma
}+\frac{b^{2}}{2}\nabla_{\beta}F^{\alpha\gamma}\nabla^{\beta}F_{\alpha\gamma}.
\label{L Pod curve 1}%
\end{equation}

Eq.~(\ref{Nao equi L1 e L2}) is useful to rewrite
Eq.~(\ref{L Pod curve 1}) as
\begin{eqnarray}
\mathcal{L}_{m} =&-&\frac{1}{4}F^{\alpha\beta}F_{\alpha\beta}+\frac{\left(
a^{2}+2b^{2}\right)  }{2}\nabla_{\beta}F^{\alpha\beta}\nabla_{\gamma}%
F_{\alpha}^{\text{ \ }\gamma}\nonumber\\
&+&b^{2}\left(  R_{\sigma\beta}F_{\text{ \ }}^{\sigma\alpha}F_{\alpha}^{\text{
\ }\beta}+R_{\alpha\sigma\beta\gamma}F^{\sigma\gamma}F^{\alpha\beta}\right).
\label{L Pod curve 2}%
\end{eqnarray}
This expression shows Podolsky electrodynamics obtained by Utiyama's
approach presents two terms with non-minimal coupling in addition to the usual term
obtained by the minimal coupling prescription in Eq.~(\ref{L Pod flat}).
It is interesting to note that non-minimally coupled terms of this form also
emerge in other contexts such as vacuum polarization in curved
background \cite{Drummond1980} and quantization of the Einstein-Maxwell theory
\cite{Deser1974}.

Hence, we can deal with Einstein-Podolsky system from two different
perspectives: we can choose minimal coupling on Eq.~(\ref{L Pod flat}) which
leads to Eq.~(\ref{L Pod curve 2}) with $b=0$, or we can choose a more general
approach where we also consider the non-minimally coupled terms preserving conditions 1 to 4 above and
characterized by $b\neq0$. The next sections are dedicated to the analyzes
of these two cases in the context of black holes with spherical symmetry.

\subsection{Field equations}

We consider Einstein-Podolsky action\footnote{We take $G=1=c$
throughout the paper.}
\begin{equation}
S=\frac{1}{16\pi}\int d^{4}x\sqrt{-g}\left[  -R+4\mathcal{L}_{m}\right]
,\label{Acao}%
\end{equation}
where $\mathcal{L}_{m}$ is given by Eq.~(\ref{L Pod curve 1}).
The corresponding field equations are obtained from the
variation of $S$ with respect to fields $A_{\mu}$ and $g^{\mu\nu}$. From
the variation with respect to $A_{\mu}$ we obtain Podolsky
equations in curved space-time:
\begin{equation}
\nabla_{\nu}\left[  F^{\mu\nu}-\left(  a^{2}+2b^{2}\right)  H^{\mu\nu}%
+2b^{2}S^{\mu\nu}\right]  =0,\label{Eq eletro two}%
\end{equation}
where $H^{\mu\nu}$ and $S^{\mu\nu}$ are antisymmetric tensors defined as
\begin{eqnarray}
H^{\mu\nu} &\equiv &\nabla^{\mu}K^{\nu}-\nabla^{\nu}K^{\mu},\label{H_munu}\\
S^{\mu\nu} &\equiv &F_{\text{ \ }}^{\mu\sigma}R_{\sigma}^{\text{ \ }\nu
}-F^{\nu\sigma}R_{\sigma}^{\text{ \ }\mu}+2R_{\text{ \ }\sigma\text{ }\beta
}^{\mu\text{ \ }\nu}F^{\beta\sigma},\label{S_munu}%
\end{eqnarray}
with
\begin{equation}
K^{\mu}\equiv\nabla_{\gamma}F^{\mu\gamma}.\label{K vector}%
\end{equation}
Variation of $S$ with respect to $g^{\mu\nu}$
leads to Einstein equations
\begin{equation}
R_{\mu\nu}-\frac{1}{2}g_{\mu\nu}R=8 \pi T_{\mu \nu} = \,\,%
8\pi\left(  T_{\mu\nu}^{M}+T_{\mu\nu}^{a}+T_{\mu\nu}^{b}\right),
\label{Eq gravitacional}%
\end{equation}
where
\begin{eqnarray}
T_{\mu\nu}^{M} &=&\frac{1}{4\pi}\left[  F_{\mu\sigma}F_{\text{ \ }\nu
}^{\sigma}+g_{\mu\nu}\frac{1}{4}F^{\alpha\beta}F_{\alpha\beta}\right],
\label{T_munu M}\\
T_{\mu\nu}^{a} &=&\frac{a^{2}}{4\pi}\left[  g_{\mu\nu}F_{\beta}^{\;\gamma
}\nabla_{\gamma}K^{\beta}+\frac{g_{\mu\nu}}{2}K^{\beta}K_{\beta}\right.
\nonumber\\
&+&\left.  2F_{(\mu}^{~\alpha}\nabla_{\nu)}K_{\alpha}-2F_{(\mu}^{~\alpha
}\nabla_{\alpha}K_{\nu)}-K_{\mu}K_{\nu}\right], \label{T_munu a}\\
T_{\mu\nu}^{b} &=&\frac{b^{2}}{2\pi}\left[  -\frac{1}{4}g_{\mu\nu}%
\nabla^{\beta}F^{\alpha\gamma}\nabla_{\beta}F_{\alpha\gamma}+F_{\text{ }(\mu
}^{\gamma}\nabla^{\beta}\nabla_{\beta}F_{\nu)\gamma}\right.  \nonumber\\
&+&\left.  F_{\gamma(\mu}\nabla_{\beta}\nabla_{\nu)}F^{\beta\gamma}%
-\nabla_{\beta}\left(  F_{\gamma}^{\text{ \ }\beta}\nabla_{(\mu}F_{\nu
)}^{\text{ \ }\gamma}\right)  \right].  \label{T_munu b}%
\end{eqnarray}
The notation $\left(  ...\right)  $ indicates symmetrization with respect to
indices $\mu\nu$. Moreover, the trace of the energy-momentum
tensor is:
\begin{eqnarray}
T&=&\frac{\left(  a^{2}+2b^{2}\right)  }{4\pi}K^{\beta}K_{\beta} \nonumber\\
&+&\frac{b^{2}}{4\pi}\left[  2\nabla_{\beta}\left(  F^{\gamma\alpha}\nabla^{\beta}%
F_{\alpha\gamma}\right)  +F^{\gamma\mu}S_{\mu\gamma}\right].
\label{Traco T_munu}%
\end{eqnarray}

Now we consider this system of field equations in the particular case of
(static) spherical symmetry. The line element can be written as
\begin{equation}
ds^{2}=e^{\nu\left(  r\right)  }dt^{2}-e^{\lambda\left(  r\right)  }%
dr^{2}-r^{2}d\theta^{2}-r^{2}\sin^{2}\theta d\phi^{2}, \label{ds}%
\end{equation}
while the field strength is given by
\begin{equation}
F_{\mu\nu}=E\left(  r\right)  \left[  \delta_{\mu}^{1}\delta_{\nu}^{0}%
-\delta_{\mu}^{0}\delta_{\nu}^{1}\right]. \label{F_munu}%
\end{equation}
So, the system is completely characterized by three functions: the
radial electric field $E\left(  r\right)  $ and the gravitational components
$\nu\left(  r\right)  $ and $\lambda\left(  r\right)  $.

In view of this parametrization we rewrite Eq.~(\ref{Eq eletro two}) as
\begin{equation}
E-\left(  a^{2}+2b^{2}\right)  \partial_{1}K_{0}+2b^{2}S_{10}=C\frac
{e^{\frac{\left(  \nu+\lambda\right)  }{2}}}{r^{2}}, \label{Eq eletro sim esf}%
\end{equation}
where $C$ is an arbitrary integration constant and
\begin{eqnarray}
K_{0}  &  =\frac{e^{\frac{\nu-\lambda}{2}}}{r^{2}}\partial_{1}\left(
r^{2}e^{-\frac{\left(  \nu+\lambda\right)  }{2}}E\right),
\label{K0 esferico}\\
S_{10}  &  =Ee^{-\lambda}\left(  \frac{\nu^{\prime}-\lambda^{\prime}}%
{r}\right). \label{S01 esferico}%
\end{eqnarray}
Prime denotes derivative with respect to the radial coordinate, e.g. $\nu' \equiv \partial_r \nu \equiv \partial_1 \nu$.
In flat space-time the spherically symmetric solution to the
electromagnetic sector Eq.~(\ref{Eq eletro sim esf}) depends on $A_{\mu}=\left(  A_{0},0,0,0\right)  $
which is a function of $x^{1}=r$ solely. It is the one due to Podolsky \cite{Podolsky1942}:
\begin{equation}
A_{0}=\frac{C_{1}}{r}-\frac{C_{2}e^{-\frac{r}{r_{p}}}}{r},
\label{A0 flat}
\end{equation}
with $r_{p}^2=a^{2}+2b^{2}$.

The non-null components of Eq.~(\ref{Eq gravitacional}) are given by
\begin{eqnarray}
e^{-\lambda}\left(  \frac{\lambda^{\prime}}{r}-\frac{1}{r^{2}}\right)
+\frac{1}{r^{2}} &  =8\pi T_{0}^{0},\label{Einstein 00}\\
-e^{-\lambda}\left(  \frac{\nu^{\prime}}{r}+\frac{1}{r^{2}}\right)  +\frac
{1}{r^{2}} &  =8\pi T_{1}^{1},\label{Einstein 11}\\
-\frac{1}{4r}e^{-\lambda}\left[  \left(  \nu^{\prime}-\lambda^{\prime}\right)
\left(  2+r\nu^{\prime}\right)  +2r\nu^{\prime\prime}\right]   &  =8\pi
T_{2}^{2},\label{Einstein 22}%
\end{eqnarray}
where
\begin{eqnarray}
T_{0}^{0} &=&-\frac{g^{00}g^{11}}{8\pi}\left\{  E\left[  E-2\left(
a^{2}+2b^{2}\right)  \partial_{1}K_{0}+4b^{2}S_{10}\right]  \right.
\nonumber\\
&+&\left.  \frac{a^{2}K_{0}^{2}}{g^{11}}+2b^{2}g^{11}\left[  \left(
\frac{K_{0}}{g^{11}}+\frac{2E}{r}\right)  ^{2}+\frac{2E^{2}}{r^{2}}\right]
\right\}, \label{T_00}\\
T_{1}^{1} &=&-\frac{g^{00}g^{11}}{8\pi}\left\{  E\left[  E-2\left(
a^{2}+2b^{2}\right)  \partial_{1}K_{0}+4b^{2}S_{10}\right]  \right.
\nonumber\\
&-&\left.  \frac{a^{2}K_{0}^{2}}{g^{11}}-2b^{2}g^{11}\left[  \left(
\frac{K_{0}}{g^{11}}+\frac{2E}{r}\right)  ^{2}+\frac{2E^{2}}{r^{2}}\right]
\right\}, \label{T_11}\\
T_{2}^{2} &=&\frac{g^{00}g^{11}}{8\pi}\left\{  E^{2}-a^{2}\left[
2E\partial_{1}K_{0}-\frac{K_{0}^{2}}{g^{11}}\right]  +\frac{2b^{2}K_{0}^{2}%
}{g^{11}}\right.  \nonumber\\
&-&\left.  4b^{2}g^{11}\left[  \left(  \frac{K_{0}}{g^{11}}+\frac{2E}%
{r}\right)  ^{2}+\frac{ES_{10}}{2g^{11}}+\frac{2E^{2}}{r^{2}}\right]  \right\}.
\label{T_22}%
\end{eqnarray}%

The requirement of spherical symmetry turns Eq.~(\ref{Traco T_munu}) for the trace of the energy-momentum tensor into:
\begin{eqnarray}
T  &=&\frac{\left(  a^{2}-2b^{2}\right)  }{4\pi}g^{00}K_{0}^{2}+\frac{b^{2}%
}{\pi}g^{00}g^{11}E\left(  \partial_{1}K_{0}-\frac{4K_{0}}{r}\right)
\nonumber\\
&+&\frac{b^{2}}{\pi}g^{00}\left(  g^{11}E\right)  ^{2}\left[  \frac{3}{2}%
\frac{\nu^{\prime}-\lambda^{\prime}}{r}-\frac{6}{r^{2}}\right].
\label{Traco esferico}%
\end{eqnarray}
This concludes our calculations of the field equation for Podolsky BH.
The solution to these equations depends on physically
meaningful boundary conditions. The first is to recover Minkowski
space-time far from the source, i.e. when $r\rightarrow \infty$ which shall be defined
as $r_{\infty}$. The second boundary is the surface of the
horizon, located by coordinate $r=r_{H}$, appearing in the black hole
solution we will study next.

\section{Analyzis of Podolsky black hole exterior solution in the case $b=0$%
\label{sec - b equal 0}}

In this section we study in detail the exterior solution of a
spherical BH in Podolsky electrodynamics particular case
corresponding to $b=0$. We show the only
non-null exterior solution is identical to Maxwell's, which means
that the BH is of the Reissner-Nordstr\"om type.

\subsection{Bekenstein's technique}

By taking $b=0$ in Eq.~(\ref{Eq eletro two}), contracting this equation with
$A_{\mu}$ and integrating the result in the 4-volume exterior
to BH, we get
\[
\int A_{\mu}\partial_{\nu}\left[  \sqrt{-g}\left(  F^{\mu\nu}-a^{2}H^{\mu\nu
}\right)  \right]  d^{4}x=0.
\]
Notice that under spherical symmetry the 4-volume is limited by the horizon $r_{H}$,
by $r_{\infty}$ ($r\rightarrow \infty$) and by the past and future infinite times
$t\rightarrow\pm\infty$. Integration by parts leads to:
\begin{equation}
I_{M}-I_{a}-\oint\sqrt{-g}A_{\mu}\left(  F^{\mu\nu}-a^{2}H^{\mu\nu}\right)
dS_{\nu}=0, \label{eq Bas}%
\end{equation}
where%
\begin{eqnarray*}
I_{M}  &=&\int\sqrt{-g}F^{\mu\nu}\partial_{\nu}A_{\mu}d^{4}x,\\
I_{a}  &=&a^{2}\int\sqrt{-g}H^{\mu\nu}\partial_{\nu}A_{\mu}
d^{4}x,
\end{eqnarray*}
with
\begin{eqnarray*}
I_{a} &=&a^{2}\int\sqrt{-g}g^{\mu\beta}g^{\nu\alpha}\left(  \partial_{\nu}A_{\mu
}\right)  \left(  \partial_{\beta}K_{\alpha}\right)  d^{4}x\\
&+&a^{2}\int K_{\beta}\partial_{\alpha}\left(  \sqrt{-g}g^{\mu\beta}%
g^{\nu\alpha}\partial_{\nu}A_{\mu}\right)  d^{4}x\\
&-&a^{2}{\oint}\sqrt{-g}g^{\mu\beta}g^{\nu\alpha}K_{\beta}\partial_{\nu
}A_{\mu}dS_{\alpha}.
\end{eqnarray*}
Therefore Eq.~(\ref{eq Bas}) is written as the sum of three
integrals:
\begin{equation}
I_{1}+I_{2}+I_{3}=0,\label{Sum I1 I2 I3}%
\end{equation}
where
\begin{eqnarray}
I_{1} &=&I_{M}-a^{2}\int K_{\beta}\partial_{\alpha}\left(  \sqrt{-g}%
g^{\mu\beta}g^{\nu\alpha}\partial_{\nu}A_{\mu}\right)  d^{4}x\nonumber\\
&-&a^{2}\int\sqrt{-g}g^{\mu\beta}g^{\nu\alpha}\left(  \partial_{\nu}A_{\mu
}\right)  \left(  \partial_{\beta}K_{\alpha}\right)  d^{4}x,\label{I1}\\
I_{2} &=&a^{2}{\oint}\sqrt{-g}g^{\mu\beta}g^{\nu\alpha}K_{\beta}\partial
_{\nu}A_{\mu}dS_{\alpha},\label{I2}\\
I_{3} &=&-{\oint}\left[  \sqrt{-g}A_{\mu}\left(  F^{\mu\nu}-a^{2}H^{\mu\nu
}\right)  \right]  dS_{\nu},\label{I3}%
\end{eqnarray}

The next step is to impose spherical symmetry: $A_{\mu
}=\left(  A_{0}(r),0,0,0\right)  $.
Consequently, the three integrals become
\begin{eqnarray}
I_{1} &=&\int\sqrt{-g}g^{00}\left[  -g^{11}E^{2}+\left(  aK_{0}\right)
^{2}\right]  d^{4}x, \label{I1 esferico}\\
I_{2} &=&a^{2}\left[  \int_{r_{H}}+\int_{r_{\infty}}\right]  \sqrt{-g}%
g^{00}g^{11}K_{0}EdS_{1}, \label{I2 esferico}\\
I_{3} &=&\left[  \int_{r_{H}}+\int_{r_{\infty}}\right]  {\frac {\sqrt{-g}A_{0}}{g_{00}g_{11}}}%
\left[  E-a^{2}\partial_{1}K_{0}\right]  dS_{1}, \label{I3 esferico}%
\end{eqnarray}
where the notation with $r_{H}$ and $r_{\infty}$
means that the integrals are performed on surfaces
of fixed $r$.
Notice that the flux integrals in
Eqs.~(\ref{I2}) and (\ref{I3}) are different from zero only on surfaces
where $r$ is constant.

Let us analyze the properties of $I_{2}$ and $I_{3}$ at
$r_{\infty}$. When $r\rightarrow\infty$ the space-time becomes flat (Minkowski)
so that
\[
\sqrt{-g}dS_{1}\approx r^{2}dS
\]
and $A_{0}$ is given by Eq.~(\ref{A0 flat}).
Using this in Eqs. (\ref{I2 esferico}) and (\ref{I3 esferico})
show the integrals over the surfaces with $r=r_\infty$ appearing in $I_{2}$ and $I_{3}$ are null.

The case for the integrals over the surfaces with $r = r_H$ is more complicated. First, we recall that the trace of the energy-momentum tensor given by
Eq.~(\ref{Traco esferico}) with $b=0$ is
\[
T=\frac{a^{2}}{4\pi}g^{00}\left(  K_{0}\right)  ^{2}.
\]
This is a scalar with physical meaning
-- it is associated with the energy of the system --, hence it
must be finite on the horizon. On the other hand, $g^{00}\left(  r_{H}\right)
\rightarrow\infty$. Consequently, $K_{0}$ must
approach zero at least at the same rate as
$\sqrt{g_{00}}$ in order to guarantee a finite value for $T$ on the horizon. In this case,
\begin{equation}
I_{2}\sim a^{2}\int_{r_{H}}Er^{2}\sin\theta\sqrt{-\frac{g_{00}}{g^{11}}}%
g^{11}g^{00}\sqrt{g_{00}}dS_{1}\sim0, \label{I2 zero}%
\end{equation}
due to the facts that the electric field is finite on
$r=r_{H}$ and $g^{11}\left(  r_{H}\right)  =0$. We conclude that the integral
$I_{2}$ is null on the horizon.

The analysis of $I_{3}$ begins by taking $b=0$ in Eq.~(\ref{Eq eletro sim esf}),
\begin{equation}
E-a^{2}\partial_{1}K_{0}=C\frac{e^{\frac{\nu+\lambda}{2}}}{r^{2}},
\label{Eq diff nao homogenea sem b}%
\end{equation}
and replacing it back into
Eq.~(\ref{I3 esferico}):
\begin{equation}
I_{3}=-C\int_{r_{H}}A_{0}\sin\theta dS_{1}. \label{I3 r horizon}%
\end{equation}
Eq.~(\ref{Eq diff nao homogenea sem b}) is a second order differential
equation for the field $E$ which may be homogeneous or non-homogeneous
according to values of the constant $C$. In flat space-time
Eq.~(\ref{Eq diff nao homogenea sem b}) becomes explicitly linear and we obtain
two homogeneous solutions, $\frac{e^{\pm\frac{r}{a}}}{r^{2}}\left(  -1\pm
\frac{r}{a}\right)  $, and one non-homogeneous solution, $\frac{C}{r^{2}}$. In curved space-time the
homogeneous solutions $E_{(h)}$ lead to $I_{3}=0$.
Then, from Eq.~(\ref{Sum I1 I2 I3}) we conclude that
\begin{equation}
I_{1}=\int\sqrt{-g}\left[  -g^{11}g^{00}E_{(h)}^{2}+g^{00}\left(  aK_{0\left(
h\right)  }\right)  ^{2}\right]  d^{4}x=0. \label{I1 homo}%
\end{equation}
Since $g_{00}>0$ and $g_{11}<0$ in the region exterior to the
horizon, each term in the square-brackets of $I_{1}$ is
positive-definite. Hence, the only possible solution
to Eq.~(\ref{I1 homo}) is:
\begin{equation}
E_{(h)}=K_{0\left(  h\right)  }=0\text{ \ \ for \ \ }r\geq r_{H}\text{.}
\label{Sol hom b igual 0}%
\end{equation}
Therefore, the existence of the horizon imposes that the
asymptotic solution ($r\gg r_{H}$) of $E_{(h)}$,
\[
E_{(h)}\simeq-C_{1}\frac{e^{-\frac{r}{a}}}{r^{2}}\left(  1+\frac{r}{a}\right),
\]
must be null, i.e. $C_{1}=0$. Notice that the demonstration fails for the
non-homogeneous solutions $E_{(Nh)}$ because $I_{3}$ is different from zero in this case.

The non-homogeneous solution of Eq.~(\ref{Eq diff nao homogenea sem b}) is:
\begin{equation}
E_{(Nh)}=C\frac{e^{\frac{\nu+\lambda}{2}}}{r^{2}}, \label{Sol Nhom b igual 0}%
\end{equation}
from which we verify that $K_{0\left(  Nh\right)  }=0$
by using Eq.~(\ref{K0 esferico}). If we replace this result in
Einstein equations Eq.~(\ref{Einstein 00}) and Eq.~(\ref{Einstein 11}) we obtain
Reissner-Nordstr\"om solution; in this case, the constant
$C$ is the electric charge.

Therefore, we conclude that the exterior solution of the Einstein-Podolsky BH
for $b=0$ is independent of the parameter $a$. This
corroborates the no-hair theorem.

\subsection{Maxwell-Proca decomposition}

For $b=0$ Podolsky field $A_{\mu}$ can be decomposed
as
\begin{equation}
A_{\mu}=A_{\mu}^{(M)}-A_{\mu}^{(P)} \label{Decomposicao}%
\end{equation}
and these components satisfy Maxwell and Proca equations:
\begin{eqnarray}
\nabla_{\nu}F^{\mu\nu(M)} &=&0, \label{Maxwell}\\
\nabla_{\nu}F^{\mu\nu(P)} &=&\frac{1}{a^{2}}A^{\mu(P)}\label{Proca}%
\end{eqnarray}
where $F_{\mu\nu}^{\left(  M\right)  }=\nabla_{\mu}A_{\nu}^{\left(  M\right)
}-\nabla_{\nu}A_{\mu}^{\left(  M\right)  }$ and similarly for $F_{\mu\nu}^{(P)}$.
Indeed, by rewriting Eq.(\ref{Eq eletro two}) as
\begin{equation}
\nabla_{\nu}\left[  F^{\mu\nu}-a^{2}\left(  \nabla^{\mu}\nabla_{\gamma}%
F^{\nu\gamma}-\nabla^{\nu}\nabla_{\gamma}F^{\mu\gamma}\right)  \right]
=0\label{Eq eletro decomposicao}%
\end{equation}
and by using Eq.~(\ref{Maxwell}) and Eq.~(\ref{Proca}) it is trivial to verify
that $F^{\mu\gamma\left(  M\right)  }$ and $F^{\mu\nu(P)}$ satisfy
Eq.~(\ref{Eq eletro decomposicao}). Besides, we can show the Podolsky
energy-momentum tensor,
\[
T_{\mu\nu}=T_{\mu\nu}^{M}+T_{\mu\nu}^{a},
\]
can also be decomposed as \cite{Zayats2013,Zayats2016}
\begin{equation}
T_{\mu\nu}=T_{\mu\nu}^{(M)}-T_{\mu\nu}^{(P)}%
,\label{Decomposicao tensor energia momento}%
\end{equation}
where $T_{\mu\nu}^{(M)}$ and $T_{\mu\nu}^{(P)}$ are the Maxwell and Proca
energy-momentum tensors, given respectively by
\begin{eqnarray}
T_{\mu\nu}^{(M)} &=&\frac{1}{4\pi}\left[  F_{\mu\sigma}^{(M)}F_{\text{ \ }%
\nu}^{\sigma(M)}+\frac{1}{4}g_{\mu\nu}F^{\alpha\beta(M)}F_{\alpha\beta}%
^{(M)}\right], \label{Tmunu Maxwell}\\
T_{\mu\nu}^{(P)} &=&\frac{1}{4\pi}\left[  F_{\mu\sigma}^{(P)}F_{\text{ \ }%
\nu}^{\sigma(P)}+\frac{1}{4}g_{\mu\nu}F^{\alpha\beta(P)}F_{\alpha\beta}%
^{(P)}\right.  \nonumber\\
&+&\left.  \frac{1}{a^{2}}\left(  A_{\mu}^{(P)}A_{\nu}^{(P)}-\frac{1}%
{2}g_{\mu\nu}A^{\beta(P)}A_{\beta}^{(P)}\right)  \right].  \label{Tmunu Proca}%
\end{eqnarray}
In this way, Einstein-Hilbert Lagrangian together with
\begin{equation}
\mathcal{L}_{m}=\mathcal{L}_{m}^{\left(  M\right)  }-\mathcal{L}_{m}^{\left(
P\right)  },\label{L Maxwell menos Proca}%
\end{equation}
where
\begin{eqnarray}
\mathcal{L}_{m}^{\left(  M\right)  } &=&-\frac{1}{4}F^{\alpha\beta
(M)}F_{\alpha\beta}^{(M)},\label{L Maxwell}\\
\mathcal{L}_{m}^{\left(  P\right)  } &=&-\frac{1}{4}F^{\alpha\beta
(P)}F_{\alpha\beta}^{(P)}+\frac{1}{2a^{2}}A^{\mu(P)}A_{\mu}^{(P)},
\label{L Proca}%
\end{eqnarray}
lead to field equations which are equivalent to Eqs.~(\ref{Eq eletro two}) and
(\ref{Eq gravitacional}) with $b=0$.

Hence, up to the negative sign in  Eq.~(\ref{L Maxwell menos Proca}),
Einstein-Podolsky and Einstein-Maxwell-Proca systems are equivalent. This
equivalence becomes clearer in the context of a static spherically symmetric BH where it has been shown independently by J.D. Bekenstein \cite{Bekenstein1972} and C. Teitelboim \cite{Teitelboin1972} that Proca fields are null in the region exterior to the
horizon. From Eq.~(\ref{Eq eletro decomposicao}) it follows that the solutions
$F_{\mu\nu}^{(P)}$ correspond exactly to the homogeneous solutions $E_{(h)}$
to Eq.~(\ref{Eq diff nao homogenea sem b}), both of which are consistent with $F_{\mu\nu}^{(P)}=E_{(h)}=0$ in the region $r\geq r_{H}$.

\section{Analyzis of Podolsky black hole exterior solution with $b\neq0$}

In this section we will study the exterior solution of a spherical BH in the presence of Podolsky electrodynamics considering parameter $b\neq0$. Analogously to Section \ref{sec - b equal 0}, we contract Eq.~(\ref{Eq eletro two}) with $A_{\mu}$ and integrate over the region exterior to the BH's horizon. This leads to
\begin{equation}
I_{1b}+I_{2b}+I_{3b}=0, \label{Sum I1b I2b I3b}%
\end{equation}
where
\begin{eqnarray}
I_{1b} &=&\int\sqrt{-g}g^{00}\left[  -g^{11}E^{2}+\left(  a^{2}%
+2b^{2}\right) K_{0}^{2}\right.  \nonumber\\
&+&\left.  2b^{2}\left(  g^{11}E\right)  ^{2}\left(  \frac{\nu^{\prime
}-\lambda^{\prime}}{r}\right)  \right]  d^{4}x,\label{I1b}\\
I_{2b} &=&\left(  a^{2}+2b^{2}\right)  \int_{r_{H}}\sqrt{-g}g^{00}g^{11}%
K_{0}EdS_{1},\label{I2b}\\
I_{3b} &=&\int_{r_{H}}\frac{\sqrt{-g}A_{0}}{g_{00}g_{11}}\left[  E-\left(
a^{2}+2b^{2}\right)  \partial_{1}K_{0}\right.  \nonumber\\
  &+&\left.  2b^{2}S_{10}\right]  dS_{1}.\label{I3b}%
\end{eqnarray}

Subtracting Eq.~(\ref{Einstein 11}) from Eq.~(\ref{Einstein 00}) implies in:
\begin{eqnarray}
\frac{\lambda^{\prime}+\nu^{\prime}}{2r}  &=&g^{00}g^{11}\left[  a^{2}\left(
\frac{K_{0}}{g^{11}}\right)  ^{2}\right.  \nonumber\\
&+&\left.  2b^{2}\left(  \left(  \frac{K_{0}}{g^{11}}+\frac{2E}{r}\right)
^{2}+\frac{2E^{2}}{r^{2}}\right)  \right].  \label{nu linha mais lambda linha}%
\end{eqnarray}
This result is then used to rewrite Eq.~(\ref{I1b}) as
\begin{eqnarray}
I_{1b} &=&\int\sqrt{-g}g^{00}\left(  -g^{11}\right)  ^{3}\left\{  \left(
\frac{E}{g^{11}}\right)  ^{2}+\frac{\left(  a^{2}+2b^{2}\right)  }{\left(
-g^{11}\right)  ^{3}}K_{0}^{2}\right.  \nonumber\\
&-&\left.  \frac{4b^{2}}{r}\frac{g_{00}^{\prime}}{g^{11}g_{00}}E^{2}%
+4a^{2}b^{2}\frac{g^{00}}{\left(  g^{11}\right)  ^{2}}rE^{2}K_{0}^{2}\right.
\nonumber\\
&+&\left.  8b^{4}g^{00}E^{2}r\left[  \left(  \frac{K_{0}}{g^{11}}+\frac
{2E}{r}\right)  ^{2}+\frac{2E^{2}}{r^{2}}\right]  \right\}  d^{4}%
x.\label{I2 final de a e b dif de zero}%
\end{eqnarray}
If we assume $g_{00}^{\prime}\geq0$ in the region exterior to the horizon
then each term of $I_{1b}$ is positive-definite. From a physical point of
view, this hypothesis is the only acceptable one once $g_{00}^{\prime}<0$ is
associated to repulsive gravity. Indeed, if exists a sub-region
$r_{1}<r<r_{2}$ exterior to $r_{H}$ where $g_{00}^{\prime}<0$, then particles
moving radially with low velocities would experience a repulsive force given
by
\[
\frac{d^{2}r}{dt^{2}}\simeq-c^{2}\Gamma_{00}^{1}\Rightarrow\frac{d^{2}%
r}{dt^{2}}\simeq c^{2}g^{11}g_{00}^{\prime}.
\]
Thus, we would have a region where the particle is impelled to move away from
the origin. An additional non-physical effect appearing if $g_{00}^{\prime}<0$
is the blue-shift of a electromagnetic wave emitted at $r_{1}$ and
detected at $r_{2}$.

The next step is to show under which conditions the integrals $I_{2b}$ and
$I_{3b}$ are null. In order to keep the trace of the energy-momentum tensor
Eq.~(\ref{Traco esferico}) evaluated at $r_{H}$ finite, $K_{0}$ must tend to
zero at least as $\sqrt{g_{00}}$. Then, from Eqs.~(\ref{I2b}) and
(\ref{I2 zero}) we have $I_{2b}\sim I_{2}\sim0$. On the other hand, field equation (\ref{Eq eletro sim esf}) may be used to cast integral $I_{3b}$ in the form:
\[
I_{3b}=-C\int_{r_{H}}A_{0}\sin\theta dS_{1}.
\]
By arguments identical to those in Section \ref{sec - b equal 0}, we see that
the homogeneous solutions $E_{(h)}$ of Eq.~(\ref{Eq eletro sim esf})
($C=0$) imply $I_{3b}=0$. In this case, under the hypothesis $g_{00}^{\prime
}\geq0$ at $r\geq r_{H}$ we conclude from Eq.~(\ref{I2 final de a e b dif de zero}) that
\begin{equation}
E_{(h)}=0\text{ \ \ for \ \ }r\geq r_{H}. \label{Sol hom b diff 0}%
\end{equation}
Hence, the only solution of Eq.~(\ref{Eq eletro sim esf}) that could possibly
be non-null is the non-homogeneous solution $E_{(Nh)}$, whose asymptotic
behavior ($r\gg r_{H}$) is of the type $C/r^{2}$. It is worth noticing that
nothing guarantees the existence of an $E_{(Nh)}$ that is consistent with the boundary
conditions imposed by a horizon. In the next section we will give an argument
contrary to the existence of a non-null $E_{(Nh)}$.

\section{Energy conditions}

In this section we will analyze Podolsky BH in the light of the null energy
condition and the weak energy condition (NEC and WEC, respectively). In particular, it is shown the
only non-trivial solution exterior to the BH horizon which does not violate NEC
and WEC is the non-homogeneous solution $E_{\left(  Nh\right)  }$ obtained
with $b=0$.

We state that the energy-momentum tensor $T_{\mu\nu}$ respects the null (weak)
energy condition if the inequality
\begin{equation}
T_{\mu\nu}k^{\mu}k^{\nu}\geq0 \label{cond energia}%
\end{equation}
holds for every null (timelike) vector $k^{\mu}$ \cite{HawEllis,Wald}. For the
particular case of a diagonal $T_{\nu}^{\mu}$, the energy conditions are
simply
\begin{equation}
\rho+p_{i}\geq0\text{ \ with \ }i=1,2,3, \label{NEC}%
\end{equation}
where $\rho\equiv T_{0}^{0}$ is the energy density and $p_{1}\equiv-T_{1}^{1}%
$, $p_{2}\equiv-T_{2}^{2}$ e $p_{3}\equiv-T_{3}^{3}$ are the principal
pressures. The WEC is satisfied if, besides
Eq.~(\ref{NEC}), we have:
\begin{equation}
\rho\geq0. \label{positividade da energia}%
\end{equation}

Eqs.~(\ref{T_00}), (\ref{T_11}) and (\ref{Eq eletro sim esf}) make it possible to rewrite Eq.~(\ref{positividade da energia}) and part of Eq.~(\ref{NEC}) as
\begin{eqnarray}
\rho &=&-\frac{g^{00}g^{11}}{8\pi}\left\{  E\left(  \frac{2Ce^{\frac{\left(
\nu+\lambda\right)  }{2}}}{r^{2}}-E\right)  +a^{2}\frac{K_{0}^{2}}{g^{11}%
}\right.  \nonumber\\
&+&\left.  2b^{2}g^{11}\left[  \left(  \frac{K_{0}}{g^{11}}+\frac{2E}{r}\right)
^{2}+\frac{2E^{2}}{r^{2}}\right]  \right\}  \geq0, \label{Cond 1}\\
\rho+p_{1} &=&-\frac{g^{00}g^{11}}{4\pi}\left\{  a^{2}\frac{K_{0}^{2}}%
{g^{11}}\right.  \nonumber\\
&+&\left.  2b^{2}g^{11}\left(  \left(  \frac{K_{0}}{g^{11}}+\frac{2E}{r}\right)
^{2}+\frac{2E^{2}}{r^{2}}\right)  \right\} \geq0.\label{Cond 2}%
\end{eqnarray}

Eq.~(\ref{Cond 2}) is satisfied in a region exterior to the horizon under two situations only, namely:

\begin{enumerate}
\item $K_{0}=0=b$ which leads to a Maxwell-like solution;

\item $K_{0}=0=E$ which implies a null field at $r\geq r_{H}$.
\end{enumerate}

In the first case, we also have $\rho+p_{2}\geq0$ and the energy density is
given by
\[
\rho=-\frac{g^{00}g^{11}}{8\pi}E^{2},
\]
which is positive-definite. Hence, for $b=0$ we conclude that the only
solution compatible with both NEC and WEC is $E=E_{(Nh)}$ given by
Eq.~(\ref{Sol Nhom b igual 0}).

In the second situation (where $b\neq0$), the only solution satisfying NEC
and WEC is the trivial solution $E=0$. This result disfavors the existence
of a non-null solution $E_{(Nh)}$ in the region exterior to the horizon.

Finally, it is interesting to note that conditions (\ref{Cond 1}) and (\ref{Cond 2}) can be used to constrain some physical configurations even out of the context of BH. For instance, from Eq.~(\ref{Cond 1}) we see that purely homogeneous solutions (i.e., those with $C=0$) will always have negative energy density. For this reason, they are physically disfavored.

\section{Final Remarks}

In this work, we have studied black holes in the presence of a matter field
given by Podolsky electrodynamics. The paper is composed of three main parts: in the first one, we presented the generalization of Podolsky electrodynamics to curved space-time; in the second part, we analyzed static spherically symmetric solutions exterior to Podolsky BH horizon; in the third part, these solutions are scrutinized in the light of the null and weak energy conditions.

The generalization of Podolsky electrodynamics to curved space-times give rise to two possible types of Lagrangian. The first one is obtained by performing the minimal coupling prescription in Eq.~(\ref{L Pod flat}) which implies $b=0$ in Eq.~(\ref{L Pod curve 1})). The second possible Lagrangian is built from Utiyama's approach \cite{Utiyama1956} (meaning $b\neq0$ in Eq.~(\ref{L Pod curve 1})). This was shown to be equivalent to the first Lagrangian up to non-minimally coupled terms depending on the contraction of the Riemann tensor and the field strength. This study has its importance not only at the classical level but also in the quantum context. For instance, we can speculate if the replacement of Maxwell theory by Podolsky's would (or not) help to control ultraviolet 1-loop divergences that are present in the Einstein-Maxwell case \cite{Deser1974}. It is worth emphasizing that something similar happens in flat space-time where Podoslky electrodynamics guarantees the finiteness of electron self-energy and vertex correction at $1$-loop \cite{Pimentel 2011}.

The exterior solutions were analyzed for two distinct cases, namely those obtained by taking $b=0$ and $b\neq0$ in the equations of motion. The only non-trivial solution for the
electromagnetic field when $b=0$ was shown to be Maxwell's solution which leads to Reissner-Nordstr\"om BH. We also verified that the Einstein-Podolsky system can be decomposed into an Einstein-Maxwell-Proca-like system, recalling that Proca field is null on BH exterior region \cite{Bekenstein1972,Bekenstein1972 v2}.
Thus, we conclude that the no-hair theorem is satisfied when $b=0$ in two different ways. For the case where $b\neq0$, we have verified that the homogeneous (asymptotically massive) solutions $E_{(h)}$ are null in the region exterior to the BH horizon under the physical hypothesis $g_{00}^{\prime}\geq0$.

Podolsky electrodynamics preserves $U(1)$ gauge invariance. Therefore, the absence of propagation of one of the Podolsky modes in the region exterior to the horizon is directly
associated to the fact that this is a massive mode; the lack of a Podolsky propagating mode is not related to the theory's gauge invariance.

In the last part, we verified that the only exterior solution consistent with
the weak and null energy conditions is Maxwell's solution, i.e. $E_{(Nh)}$
with $b=0$. Therefore, any possible non-Maxwellian solution (a solution with
hair -- e.g. $E_{(Nh)}$ with $b\neq0$) necessarily violates NEC and WEC. Moreover, it was shown that any purely homogeneous exterior solution to Podolsky BH has a negative-definite energy density. Particularly, this can be verified for the case $b=0$ in the context of Maxwell-Proca decomposition; then, $T_{\mu\nu}^{(M)}$ is null since $E_{(Nh)}=0$, leading to $T_{\mu\nu}=-T_{\mu\nu}^{(P)}$. This decomposition shows our Proca field is a ghost field.

The conclusion is: under reasonable physical hypotheses, the static spherically symmetric Podolsky BH satisfies the no-hair theorem. However, for
$b\neq0$, solutions with hair are not mathematically excluded. In a future
work, it would be interesting to investigate if these solutions exist and what
properties they possess.

\begin{acknowledgments}
RRC is grateful to R.H. Brandenberger for the kind hospitality
extended to him at McGill and CAPES-Brazil (Grant 88881.119228/2016-01)
for financial support. BMP acknowledge CNPq-Brazil for
partial financial support. LGM is grateful to CNPq-Brazil (Grant 112861/2015-6)
for financial support. The authors thank E.M. de Morais for useful discussions during the early stages of this work.
\end{acknowledgments}

\end{document}